\documentclass[a4paper]{jpconf}
\usepackage[dvips]{graphicx}
\begin{document}

\newcommand{\GeVc}    {\mbox{$ {\mathrm{GeV}}/c                            $}}
\newcommand{\GeVcc}{\mbox{${\rm GeV/c^2}$}}
\newcommand{\MeVc}    {\mbox{$ {\mathrm{MeV}}/c                            $}}
\newcommand{\hetrois}    {\mbox{$ ^{3}{\mathrm{He}}                            $}}
\newcommand{\hetro}    {\mbox{$ ^{3}{\mathrm{He}}                            $}}
\newcommand{\xe}    {\mbox{$ ^{129}{\mathrm{Xe}}                            $}}
\newcommand{\ger}    {\mbox{$ ^{73}{\mathrm{Ge}}                            $}}
\newcommand{\al}    {\mbox{$ ^{27}{\mathrm{Al}}                            $}}
\newcommand{\tritium}    {\mbox{$ ^{3}{\mathrm{H}}                            $}}
\newcommand{\hequatre}    {\mbox{$ ^{4}{\mathrm{He}}                            $}}
\newcommand{\he}{$^4$He }
\newcommand{\hee}{$^4$He}
\newcommand{\het}{$^3$He }
\newcommand{\hett}{$^3$He}
\newcommand{\fe}{$^{55}$Fe }
\newcommand{\alu}{$^{27}$Al }
\newcommand{\cm}{$^{244}$Cm }
\newcommand{\iso}{C$_4$H$_{10}$ }
\newcommand{\isoo}{C$_4$H$_{10}$}
\newcommand{\nit}{N$_4$S$_3$ }
\newcommand{\neut}{$\tilde{\chi}^0$}
\newcommand{\neutt}{$\tilde{\chi}$}

\def\Journal#1#2#3#4{{#1} {\bf #2}, #3 (#4)}
\def\NCA{\em Nuovo Cimento}
\def\NIMA#1#2#3{{\rm Nucl.~Instr.~and~Meth.} {\bf{A#1}} (#2) #3}
\def\NIM#1#2#3{{\rm Nucl.~Instr.~and~Meth.} {\bf{#1}} (#2) #3}
\def\NPB{{\em Nucl. Phys.} B}
\def\PLB{{\em Phys. Lett.}  B}

\def\PRA#1#2#3{{\rm Phys. Rev.} {\bf{A#1}} (#2) #3}
\def\PRB#1#2#3{{\rm Phys. Rev.} {\bf{B#1}} (#2) #3}
\def\PRC#1#2#3{{\rm Phys. Rev.} {\bf{C#1}} (#2) #3}
\def\PRD#1#2#3{{\rm Phys. Rev.} {\bf{D#1}} (#2) #3}

\def\JHEP#1#2#3{{\rm JHEP} {\bf{#1}} (#2) #3}
\def\ZPC{{\em Z. Phys.} C}
\def\PRL#1#2#3{{\rm Phys.~Rev.~Lett.} {\bf{#1}} (#2) #3}
\def\PLB#1#2#3{{\rm Phys.~Lett.} {\bf{B#1}} (#2) #3}
\def\APP#1#2#3{{\rm Astropart.~Phys.} {\bf{B#1}} (#2) #3}
\def\APJ#1#2#3{{\rm Astrophys.~J.} {\bf{#1}} (#2) #3}
\def\APJS#1#2#3{{\rm Astrophys.~J.~Suppl.} {\bf{#1}} (#2) #3}
\def\AA#1#2#3{{\rm Astron. \& Astrophys.} {\bf{#1}} (#2) #3}
\def\JCAP#1#2#3{{\rm JCAP} {\bf{#1}} (#2) #3}

\title{Low energy measurements with Helium Micromegas~$\mu$TPC}
\author{O. Guillaudin$^1$, F. Mayet$^1$, C. Grignon$^1$, C. Koumeir$^1$, D. Santos$^1$, P. Colas$^2$, I. Giomataris$^2$}
\address{$^1$ LPSC, Universit\'e Joseph Fourier Grenoble 1,
 CNRS/IN2P3, Institut Polytechnique de Grenoble, 53 avenue des Martyrs, 38026 Grenoble, France}
\address{$^2$ IRFU/DSM/CEA, CE Saclay, 91191 Gif-sur-Yvette cedex, France}

\ead{olivier.guillaudin@lpsc.in2p3.fr}

\begin{abstract}
The measurement of the ionization produced by particles in a medium presents a great interest 
in several fields from metrology to particule physics and cosmology. The ionization quenching factor 
is defined as the fraction of energy released by ionisation by a recoil in a medium    compared with its 
  kinetic energy. At low energy, in the range of a few keV, the ionization falls rapidly and 
systematic measurement are needed. We have developped an experimental setup devoted to the measurement of low energy  (keV) 
ionization quenching factor for the MIMAC project. The ionization produced in the gas has 
been measured with a Micromegas detector filled with Helium gas mixture.
\end{abstract}

\section{Introduction}
A nuclear recoil moving at very low energies can be used to detect rare events such as neutrino coherent interactions or 
non-baryonic dark matter signatures. The direct detection of these non-baryonic particules is based on the detection 
of nuclear recoils coming from elastic colisions on different targets. Ionization is one of the most important channel to detect such nuclear recoils. The ionization quenching factor is defined as the fraction of the kinetic energy released through ionization by a recoil in a medium. In the last decades, an important effort  has been made to measure the IQF in different materials : 
gases \cite{H}, solids \cite{Ge,Si} and liquids \cite{Xe}, using different techniques. However, in the low energy range, 
the measurements are rare or absent for many targets due to ionization threshold of detectors and experiment contraints.\\
We have developped an experimental setup devoted to the measurement of low energy  (a few keV) 
ionization quenching factor for the MIMAC project \cite{mayet-susy,moulin}.

The measurements reported in this work have been done with a micromegas detector filled with a 
gas mixture of 95\% of \he and 5\% of isobutane (\isoo) at different pressures. 
The Micromegas detector was calibrated using X-rays produced by fluorescence on different targets. 
To produce nucleus moving with a controlled energy in the detection volume, we have developped calibrated 
Electron Cyclotron Resonance Ion Source (ECRIS) \cite{geller} which produced ions  with energy from a 
fraction of one keV up to 50 keV. The ECRIS source was coupled to a Micromegas detector \cite{mayetparis}.

\section{Experimental setup and detector calibration}
The micromegas used in this work was a standard Bulk Micromegas \cite{bulk}, 
in which the mesh and the anode are built and integrated with a fixed  128 $\mu m$ gap.  
The gap between the copper anode and stainless steel mesh is maintained by  cylindrical pilars every 3 mm. 
This gap is well adapted for working gas pressure between 350 and 1300 mbar. 
The drift distance between the cathode and the mesh was 3 cm, large enough to include the tracks of nuclei of energies 
up to 50 keV at 350 mbar. The active area was $\rm 100 \times  100 \ mm^2$.

We used an individual power supply for each electrode (mesh and anode) with
cathode drift to the ground, allowing to independently vary the different
fields. The NIM High Voltage module (ISEG 223M) presents a high stability and
high precision of the output voltage setting and measurement (about 10 mV). The
setting parameters, the voltage and the current controle are performed by remote
control via a LabView interface. Typical applied field were $\rm 100 \ V/cm$ for
the drift and a voltage of 450 V for the avalanche. With these values, an optimum collection efficiency is achieved and the electron drift velocity is comparatively low.

The electronic system is composed of a dedicated charge pre-amplifier and of a commercial spectrocopy amplifier. Due to the low drift velocity of electron in such a low drift field ($\rm 100 \ V/cm$), the integration time constant of this pre-amplifier should be large enough to assure a total colection charge and 
an output proportionnal to the total energy of the event. The calibration constant for the preamplifier is measured by applying test pulse on a test capacity. A PCI-bus multichannel analyzers with internal 8k ADC is used to convert incoming signals.

For each measurement, the gas vessel is previously pumped using a turbomolecular pump and a dry 
primary vacuum pump. Then the detector is filled with gas mixture at the working pressure and operates with a circulating and controlled
flux. To correct this low gain variation, a calibration is performed just before each energy measurement with the ion source.

\begin{figure}[t]
\begin{center}
\includegraphics[scale=0.4]{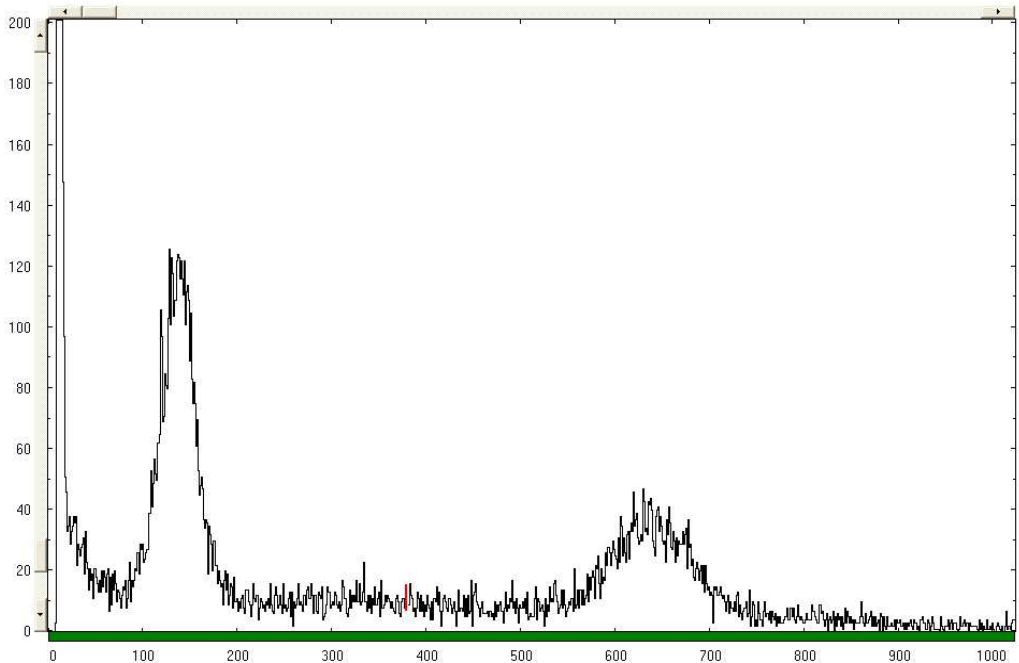}
\caption{Response of \he + 5\% \iso Micromegas to Aluminium (1.486 keV) and Iron 
(6.4 keV) fluorescence X-rays at 1000 mbar, with  a resolution (FWHM) of 29,4 \% and 15,3 \% respectively.}
\label{fg.calibration}
\end{center}
\end{figure}

Due to the high transparency of \he gas for X-rays energy above 2 keV, the detector shows a 
very low efficiency to X-ray from a \fe source used for calibration purpose, especially at 
low pressure. So, to avoid using high intensity radioactive source and to calibrate in a realistic time, 
we have developped a high intensity multi energies X-ray system based on the production of fluoresence photons 
on different high purity targets. The system used a pulsed miniature X-rays generator inside the gas vessel 
with maximum flux equivalent to a 2 mCi radioactive source and the following targets with the corresponding X-ray 
energies :  Aluminium (1.486 keV), Titanium (4.504 keV), Iron (6.4 keV), Copper (8.1 keV).
Figure \ref{fg.calibration} shows a typical spectrum for Aluminium (1.486 keV) and Iron (6.4 keV) targets 
with  a resolution (FWHM) of 29,4 \% and 15,3 \% respectively. The energy threshold is close to 300 eV.

\section{The ion source}
\begin{figure}[t]
\begin{center}
\caption{Electron Cyclotron Resonance Ion Source (ECRIS) and the 45 degree spectrometer (yellow part) coupled to the Micromegas detector.}
\label{fg.source}
\end{center}
\end{figure}

The ECRI ion source is a plasma device designed and built by the LPSC SSI team to provide charged ions at low velocities from 
a fraction of a keV  to 50 keV. One of the important features of this facility is the availability of very low and stable 
currents of a few picoamp. In the present configuration, the beam is adapted to the 45 degree spectrometer by an Einzel lens which has been tuned with the faraday cup. A micrometer aperture
($1 \ \mu m $ in diameter) selects a part of this beam which is injected in the Micromegas detector and produced about 
25 ions per second. According to the gas injected is the plasma chamber, the source is able to 
produce many type of beams : proton, $^3$He, $^4$He, $^{19}$F.\\
To calibrate the ion source, a 50 nm thick Silicon Nitride membrane (Si$_3$N$_4$) was used as interface between the ion source and the gas
chamber. A time of flight measurement has been performed under vacuum, using two channeltrons. One of them detecting the low energy electrons extracted from the Si$_3$N$_4$ menbrane by the ions and the other one detecting the ions for 6 different known positions. This setup allows to measure the energies of the ions just after the foil. With this configuration a first energy measurement with micromegas was performed.
In such a way, we could verify that the energy measured by TOF were the same as those indicated by the extraction potentiel  values in kV for the 1+ charge state ions with the $1 \ \mu m $ diameter aperture interface.

\section{Low energy ion measurements}
\begin{figure}[t]
\begin{center}
\includegraphics[scale=0.4,angle=270]{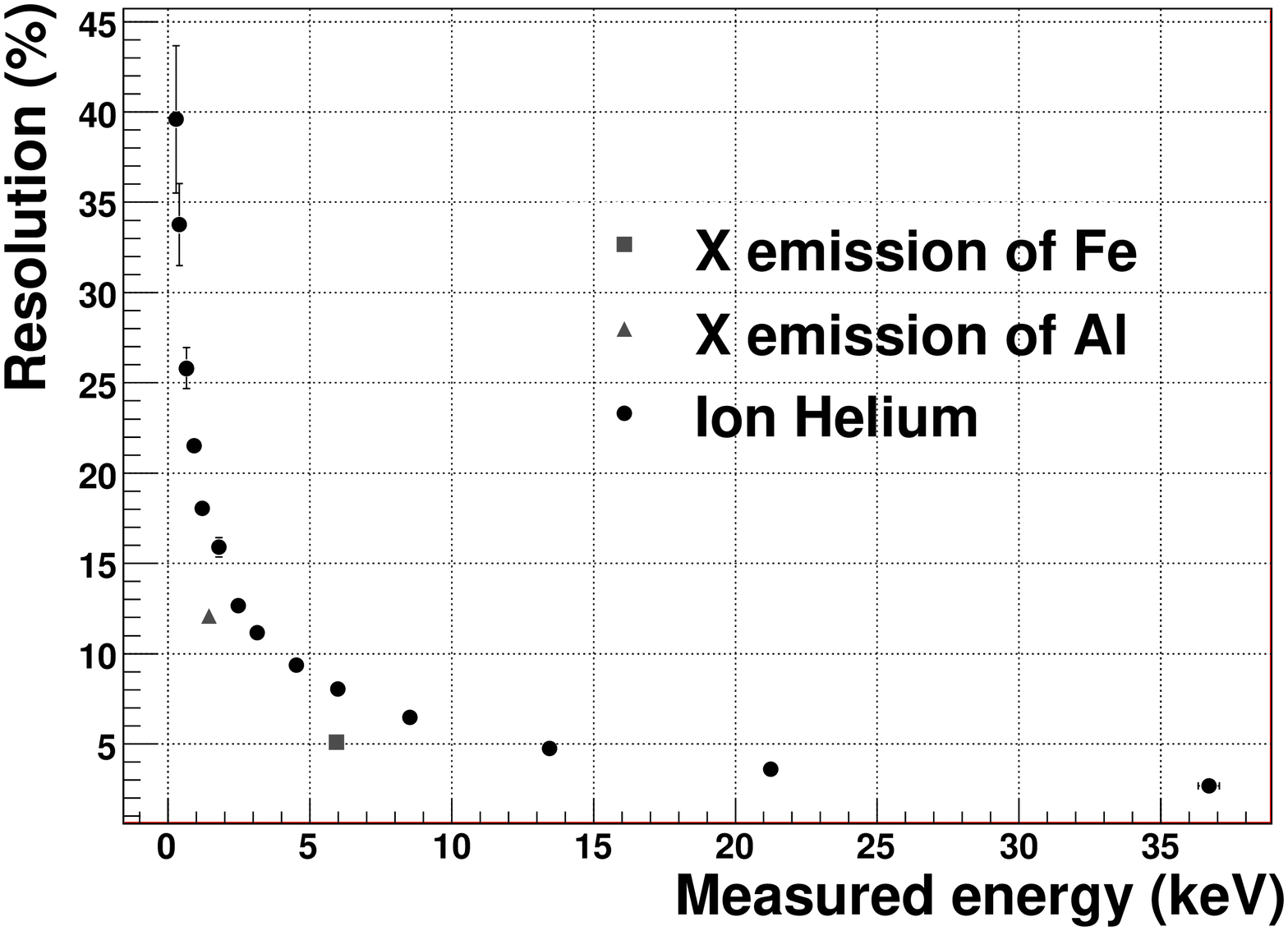}
\caption{Energy resolution for X-rays and \he ions in a \he + 5\% \iso Micromegas detector at 700 mbar.}
\label{fg.resol}
\end{center}
\end{figure}
Using the photoelectron produced by fluorescence X-rays as calibration, we can measure ionization energies produced by the recoils. The ionization quenching factor of a recoil will be the ratio between this energy and 
its   kinetic energy. 
The measurement reported \cite{prl} have been focused on the low energy \he IQF \cite{mayetparis}.\\ At the same time, it was possible to extract the Micromegas energy resolution for a wide range of incoming ion energies. Ionization energies have measured for ions with kinetic energy down to 1 keV. Figure \ref{fg.resol} presents the evolution of the resolution (FWHM) at 700 mbar for measured ionization energies of \he ions from 300 eV to 35 keV. We can note that the resolution 
is slightly better for X-rays. At low energy, down to 1 keV, the energy resolution is about 50\%, but this does not affect 
the number of expected events for Dark Matter search \cite{mayetparis}.\\
Using the possibility to produce recoils of different states of charge (0,+1,+2 for \he) with the same kinetic energy, 
it is also possible to confirm that the ionization energy does not depend of the state of charge of the recoil. 
This was clearly demonstrated with \he recoil with 50 keV of kinetic energy. Measurement with neutral 
recoils have been obtained with the Si$_3$N$_4$ window because ions are neutralized when passing 
through the membrane. Moreover, this experimental setup allow to measure the quenching factor for other 
different recoils wich could be used for dark matter search such as : $^3$He in $^3$He or $^{19}$F in CF$_4$.\\
Concerning $^3$He, which is a rather expensive gas, a clean close loop gas system is under construction to allow gas recovery and storage after the measurements.

\section{Conclusion}
In summary, we have demonstrated the possibility to measure ionization energy for \he recoils down to 1 keV with a compatible energy resolution for Dark Matter search. For the next step, this measurement will be extended to low gas pressure and to different recoil such as $^3$He of 
$^{19}$F. This measurement is particularly important to better understand the ionization response of gas detectors.

\end{document}